\documentclass[%
 reprint,
superscriptaddress,
 amsmath,amssymb,
 aps,
]{revtex4-2}
  
\usepackage{graphicx}
\usepackage{dcolumn}
\usepackage{bm}

\usepackage[caption=false]{subfig}

\usepackage{url}
\usepackage{braket}
\usepackage{hyperref}
\hypersetup{
  colorlinks   = true,    
  urlcolor     = blue,    
  linkcolor    = blue,    
  citecolor    = blue,    
}

\usepackage{amsmath} 
\usepackage{siunitx}
\usepackage{placeins}

\begin{document}

\preprint{APS/123-QED}

\title{ Pump-induced terahertz anisotropy in bilayer graphene } 

\author{Angelika Seidl}
\email{a.seidl@hzdr.de}
\affiliation{Helmholtz-Zentrum Dresden-Rossendorf, 01328 Dresden, Germany}
\affiliation{Technische Universität Dresden, 01062 Dresden, Germany}
\author{Roozbeh Anvari}
\affiliation{Queen's University, Kingston K7L 3N6, Ontario, Canada}
\author{Marc M. Dignam}
\affiliation{Queen's University, Kingston K7L 3N6, Ontario, Canada}
\author{Peter Richter}
\affiliation{Technische Universität Chemnitz, 09111 Chemnitz, Germany}
\author{Thomas Seyller}
\affiliation{Technische Universität Chemnitz, 09111 Chemnitz, Germany}
\author{Harald Schneider}
\affiliation{Helmholtz-Zentrum Dresden-Rossendorf, 01328 Dresden, Germany}
\author{Manfred Helm}
\affiliation{Helmholtz-Zentrum Dresden-Rossendorf, 01328 Dresden, Germany}
\affiliation{Technische Universität Dresden, 01062 Dresden, Germany}
\author{Stephan Winnerl}
\affiliation{Helmholtz-Zentrum Dresden-Rossendorf, 01328 Dresden, Germany}
\date{\today} 

\begin{abstract}
   We investigate the intraband nonlinear dynamics in doped bilayer graphene in the presence of strong, linearly-polarized, in-plane terahertz fields. We perform degenerate pump-probe experiments with \SI{3.4}{THz} fields on doped bilayer graphene at low temperature (\SI{12}{K}) and find that when the pump is co-polarized with the probe beam, the differential pump-probe signal is almost double that found in the cross-polarized case. We show that the origin of this pump-induced anisotropy is the difference in the average electron effective mass in the probe direction when carriers are displaced in $k$-space by the pump either parallel or perpendicular to the direction of the probe polarization. We model the system using both a simple  semiclassical model and a Boltzmann equation simulation of the electron dynamics with phenomenological scattering and find good qualitative agreement with experimental results.

\end{abstract}

\maketitle


\section{Introduction}

Bilayer graphene features several interesting and unusual characteristic properties, such as a $4\pi$ pseudospin rotation \cite{MacDonald2012}, a tunable band gap \cite{Zhang2009}{}, and superconductivity in “magic-angle” twisted bilayers \cite{Cao2018}{}. The maximum of the highest valence band and the minimum of the lowest conduction band touch at the K and K' points of the Brillouin zone. Both bands are symmetric to each other and approximately parabolic for low energies. At higher energies the dispersion of the bands becomes linear with a Fermi velocity similar to monolayer graphene \cite{McCann2006}{}. Applying an electric field perpendicular to the graphene layers results in the opening of a band gap \cite{Zhang2009}{}. 

Doped bilayer graphene is an attractive material for high-mobility transport applications and terahertz (THz) optoelectronics, \textit{e.g.} for analog transistors \cite{Fiori2014}{}, sensitive THz detectors \cite{Spirito2014,Qin2017}{}, THz plasmonic structures \cite{Jadidi2019}  and modulators \cite{Liu2013}{}.  While there are many studies on carrier dynamics and nonlinear effects at THz frequencies in monolayer graphene, these fundamental properties are essentially unexplored experimentally in bilayer graphene. 
In monolayer graphene, nonlinear THz transmission experiments \cite{Paul2013,Mics2015}{}, degenerate THz pump-probe experiments \cite{Winnerl2011,Hwang2013}{}, optical-pump THz-probe experiments \cite{Jnawali2013,Tielrooij2013}{} and THz-pump optical-probe experiments \cite{Tani2010,Melnikov2019}{} have all been performed, and high-harmonic generation has been demonstrated \cite{Hafez2018}{}. For a review see Refs. \cite{Hafez2020,Massicotte2021}{}. For both monolayer and bilayer graphene there are theoretical predictions of nonlinear THz effects based on the transient currents resulting from the non-parabolic band structures \cite{Mikhailov_2008,Ang2010,McGouran2016}{}. Although the intrinsic nonlinearity of monolayer graphene plays a role, many of the above-mentioned experimental phenomena can be largely explained as arising from ultrafast carrier thermalization, which results in a hot Fermi-Dirac distribution. The hot carriers result in a reduced THz conductance and cool down on a timescale of few ps \cite{Mics2015}{}. In a recent THz-pump optical-probe experiment, a small anisotropy in the hot carrier distribution of the order of $10^{-5}$ has been observed \cite{Melnikov2019}{}, which to our knowledge so far is the only anisotropic effect arising from intraband carrier dynamics. This anisotropy indicates that there exist effects beyond the isotropic hot carrier response.\\

Here we present an experimental study complemented by microscopic theory of the nonlinear THz pump-probe response of bilayer graphene. We find that, almost independent of the pump fluence, the induced transmission is about twice as large when pump and probe beam are co-polarized than when they are cross-polarized. Using a simple semiclassical theory, we demonstrate that the origin of this difference arises from the strong  $k$-vector dependence of the effective mass of the electrons in the direction of the probe polarization. 


\section{Experimental System}

In our experiment we use a 10$\times$\SI{10}{\square\milli\metre} bilayer graphene sample on SiC. 
The sample was fabricated as follows: In a first step, a monolayer of graphene is grown on 6H-SiC(0001) \cite{Mammadov2014}{}.
This monolayer and the underlying buffer layer are converted into a graphene bilayer via hydrogen intercalation and subsequent annealing at \SI{860}{\degreeCelsius} \cite{Riedl2009}{}. 
The spontaneous polarization of SiC induces a band gap of about 120 to \SI{150}{\milli\electronvolt} in the bilayer \cite{Ohta2006,Mammadov2017}{}. Samples grown by this method are p-type with a typical hole concentration of \SI{6.5e12}{\per\square\centi\metre} \cite{Mammadov2014}{}. The bilayer nature of our sample is confirmed by x-ray photo-electron spectroscopy (XPS) and atomic-force microscopy (AFM).\\

The free-electron laser FELBE was used as source of intense THz radiation to pump and probe the sample at \SI{3.4}{THz} (photon energy \SI{14}{\milli\electronvolt}, pulse duration \SI{13}{\pico\second}, repetition rate \SI{13}{MHz}). Since the photon energy is much smaller than the Fermi energy and the band gap, interband transitions are inhibited for moderate THz fields and the measured signals stem from intraband transitions. The pump beam is normally incident on the sample and the probe beam offset by a small angle. We apply spatial filtering to suppress scattered pump radiation. Both beams were focused using a parabolic mirror (focal length \SI{80}{\milli\metre}) yielding a pump beam diameter of \SI{500}{\micro\metre} (FWHM) on the sample. In both the pump and the probe beam path, wire-grid polarizers were used to switch between horizontal and vertical polarization. To this end, two polarizers were included in each path, the first one serving for control of the power, and the second one to determine the polarization. The time delay between pump and probe pulse is changed using a delay stage moving with constant velocity (measurement "on-the-fly") to average out the contribution arising from phase oscillation between pump and probe pulse.\\

The transmission change for the probe beam was detected with a He-cooled bolometer. Chopping the pump beam and processing the signal with a lock-in amplifier, the differential transmission signals were recorded. The integration time of the lock-in amplifier was \SI{1}{\second}, longer than the time needed to scan over a few interference fringes. All measurements were performed at low temperature (\SI{12}{K}). \\


\section{Experimental Results}

Measuring all four combinations of horizontally and vertically polarized pump and probe beams, we confirmed that the anisotropy of the pump-probe signal depends on the relative  orientation of the beams but not on their absolute orientation (see Appendix A). This excludes polarization effects in the setup as well as polarization effects depending on the crystallographic axis of the sample. For the results presented here we kept the probe beam horizontally polarized and varied only the polarization of the pump beam. \\

\begin{figure}
 \centering
 \subfloat[\label{fig_1_a}]
   \centering
   \includegraphics[width=\columnwidth]{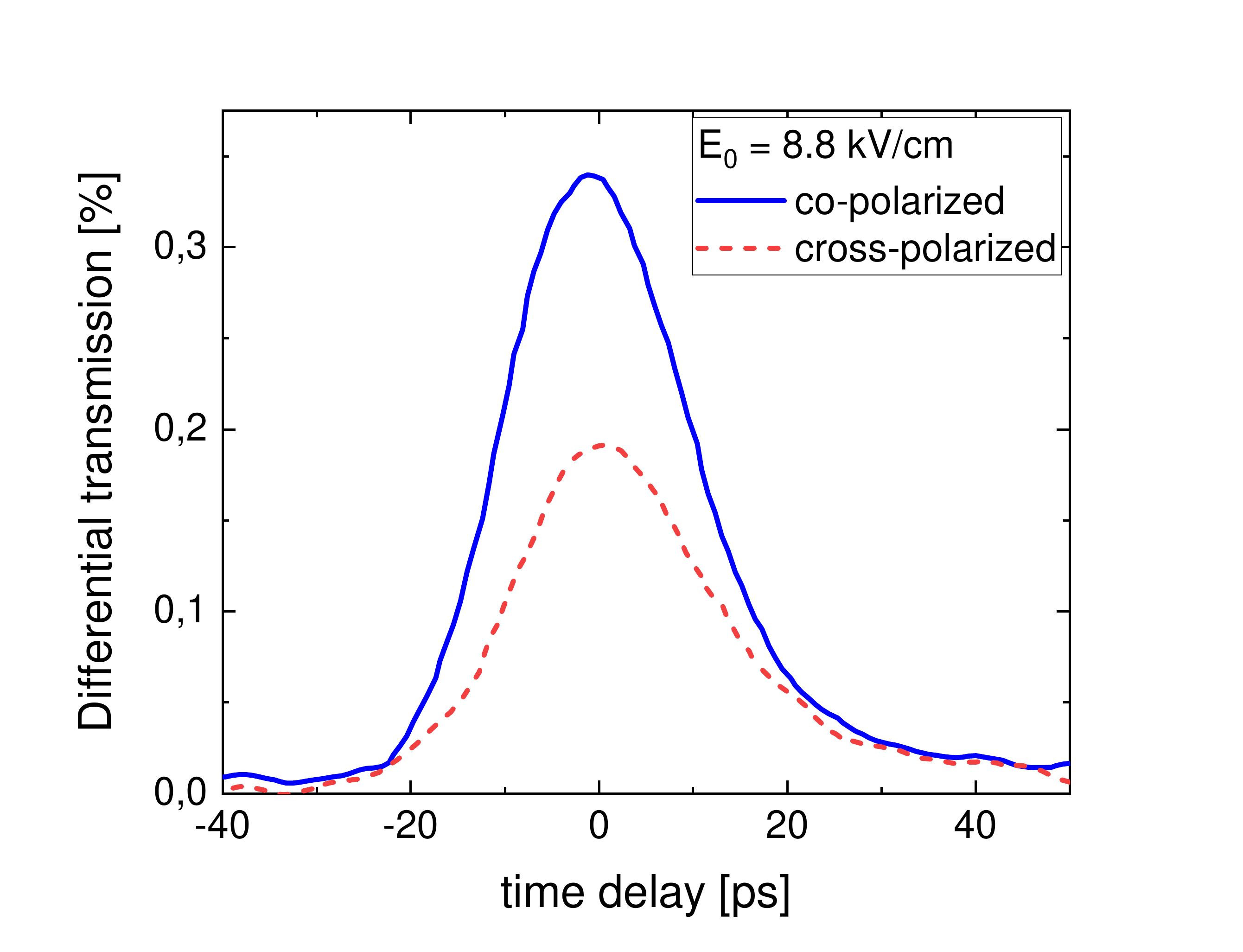}
\hfill
\subfloat[\label{fig_1_b}]
   \centering
   \includegraphics[width=\columnwidth]{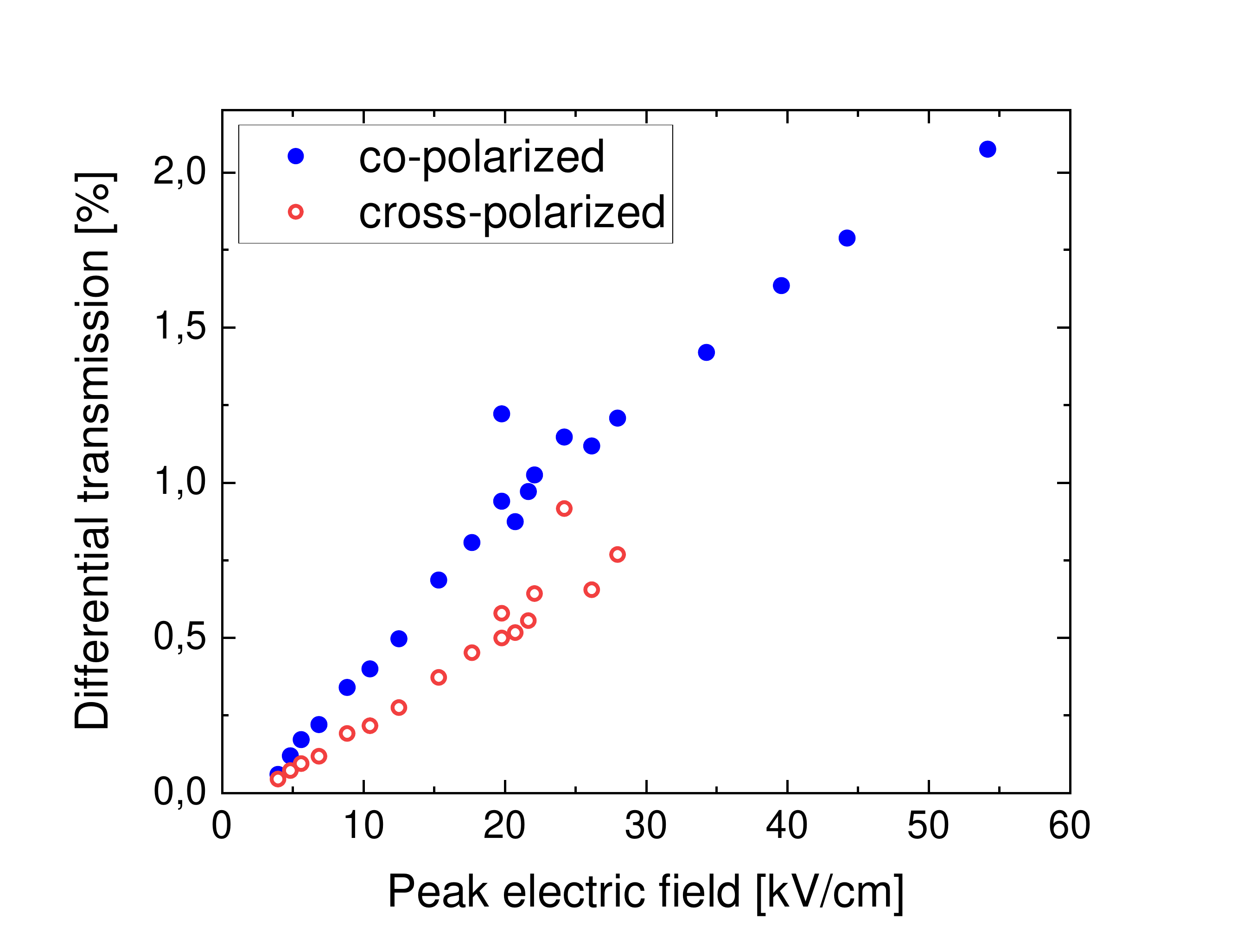}
\caption{ (a) Differential transmission signals for co- and cross-polarized pump and probe beams as a function of probe time delay for a pump field amplitude of \SI{8.8}{\kilo\volt\per\centi\metre}. (b) Amplitude of the differential transmission signals at zero probe delay as a function of the pump field amplitude.}
  \label{fig_1}
\end{figure}

In Fig. \ref{fig_1_a}, we plot the induced differential transmission signals for a co-polarized and cross-polarized pump as a function of the probe time delay for a pump field amplitude of \SI{8.8}{\kilo\volt\per\centi\metre}. As can be seen, at zero delay, the signal for the co-polarized case is about 1.8 times larger than in the cross-polarized case. The positive signals, corresponding to an increase in transmission - due to a reduction in the intraband conductivity - upon pumping, is similar to the hot-carrier response of monolayer graphene described in the introduction. The difference for the two polarization configurations, however, indicates that the carrier distribution during the pump excitation is strongly anisotropic. The decay of the signal is for both polarization configurations only slightly longer than the rising edge. This indicates that the main part of the signal stems from ultrafast processes during the \SI{13}{\pico\second} long pulse. The ultrafast dynamics contains both coherent electron dynamics and carriers that have undergone momentum scattering in the time range of 30 to \SI{190}{\femto\second} \cite{Hong2009,Mics2015,JKO2017}{}. Furthermore, there is a decay on the scale of a few picoseconds related to carrier cooling \cite{Hwang2013,Jnawali2013}{}. This cooling is known to arise from either supercollisions \cite{Song2012} or cooling of the high-energetic tail of the hot electron distribution by optical phonons \cite{Winnerl2011,Pogna2021} or both.

For both polarization configurations, the pump power was increased from the lowest value that had a significant signal-to-noise ratio up to the maximum available power. From the spot diameter measured with a pyroelectric camera and the pulse duration, which can be deduced from the measured spectrum of the Fourier limited FELBE pulses \cite{Michel2004First-8875,8662700} and the risetime of the pump-probe signals, we calculated the peak field corresponding to the pump power. As depicted in Fig. \ref{fig_1_b}, over this range of pump field amplitudes, the differential transmission signal amplitude increases approximately linearly with the pump electric field for both co- and cross-polarized configurations, implying a square-root dependence on the applied pump power. The anisotropy factor, defined as the ratio of co-polarized to cross-polarized signal maxima, decreases slightly over the whole measurement range.\\
We note that the local temperature of the graphene recovers to \SI{12}{K} in between two FEL pulses. This can be deduced from a comparison of our experimental parameters with the calculations by Xu and Buehler \cite{Xu2012} (for the temperature dependence of the graphene heat capacity, see \cite{Sang2019}{}).

\FloatBarrier
\section{Theory and simulations}

In order to obtain an intuitive understanding of the experimental results, we first provide a simple semiclassical description of the response. We follow this up with a full simulation of the dynamics. \\
\subsection{Semiclassical model}
The basic semiclassical model we employ is similar to that of Mikhailov \cite{Mikhailov2007}{}, but applied to a pump-probe experiment with the inclusion of phenomenological scattering. We start with the semiclassical equation for the average of the electron wave vector,
\begin{equation}
\label{eq_Boltz}
\frac{d\mathbf{k}\left(  t\right)  }{dt}=-\frac{e}{\hbar}\mathbf{E}_t\left(
t\right)  -\frac{\mathbf{k}\left(  t\right)  }{\tau},
\end{equation}
where $e$ is the elementary charge, $\tau$ is a phenomenological scattering time, and $\mathbf{E}_t\left(  t\right)  =\mathbf{E}_{to}\cos(\omega t)$ is the THz electric field at the graphene (\textit{i.e.} the transmitted field). For a more intuitive understanding, we consider electrons in the conduction band, which are fully symmetric to holes in the valence band for bilayer graphene in the range of a few hundred meV.  \\ 

The current density is given by 
\begin{equation}
    \label{eq_current}
\mathbf{J}\left(  t\right)  =\frac{-2e}{A}{\displaystyle\sum\limits_{\mathbf{k}}}
\rho_{cc}\left(  \mathbf{k},t\right)
\mathbf{v}\left(  \mathbf{k} \right),
\end{equation}
where $A$ is the area of the graphene, the factor of 2 accounts for the two spins,
$\rho_{cc}\left(  \mathbf{k},t\right)$ is the conduction band element for the density matrix, and $\mathbf{v}\left(  \mathbf{k}\right)  \equiv \boldsymbol{\nabla
}_{\mathbf{k}}E_{c}\left(  \mathbf{k}\right) /\hbar$ is the carrier velocity. From Eq. (\ref{eq_Boltz}), we see that the electrons on average will all move in phase and the occupation probability is one for all states inside the disk of radius $k_F$ centered at $\mathbf{k}_{c}\left(  t\right)$, which is found by solving Eq. (\ref{eq_Boltz}) with $\mathbf{k}_{c}\left(  0\right)$ =0. Now let us assume that at $t=t_{o}$, the carriers are
in the ground state at zero temperature, such that all states are occupied
that have an energy less than the Fermi energy $E_{F}$. Since close to the
Dirac points, the bands are rotationally symmetric, this also means that all
states that have a k-vector less than $k_{F}$ relative to the given Dirac point are occupied, where $E_{F}=E_{c}\left(  k_{F}\right)$.
Considering the average effects of scattering, 
all of the electrons will move in phase together and the
occupied states will simply be in a disk of radius $k_{F}$ with the
center given by
\begin{equation}
\mathbf{k}_{c}\left(  t\right)  =-\frac{e\mathbf{E}_{to}\tau\cos\left(  \omega
t-\phi\right)  }{\hbar\sqrt{1+\omega^{2}\tau^{2}}}.
\end{equation}
Thus, moving to a continuum of k-states and accounting for the two Dirac valleys,
the current density is given by
\begin{equation}
\mathbf{J}\left(  t\right)   =-\frac{4e}{\hbar\left(  2\pi\right)  ^{2}}%
{\displaystyle\int} d^{2}\boldsymbol{\nabla
}_{\mathbf{k}}E_{c}\left(  \mathbf{k+k}_{c}\left(  t\right)
\right)  \Theta\left[  k_{F}-k\right],
\end{equation}
where $k=|\mathbf{k}|$ and $\Theta\left(  k\right)  $ is the Heaviside step function. \\

Let us consider the pump-probe experiment where there is no delay between the pump and probe fields. We take the electric field amplitude at the graphene to be given by%
\begin{equation}
\mathbf{E}_{to}=\frac{1}{2}\left[  E_{tp}\widehat{\mathbf{e}}_{p}%
+E_{ts}\widehat{\mathbf{e}}_{x}\right]  ,
\end{equation}
where $E_{tp}$ is the pump field amplitude and $E_{ts}$ is the probe (signal)
field amplitude at the graphene. Note that both are real and positive
because they are in phase. For the parallel pump case, $\widehat{\mathbf{e}%
}_{p}=\widehat{\mathbf{e}}_{x}$, while for the perpendicular pump case,
$\widehat{\mathbf{e}}_{p}=\widehat{\mathbf{e}}_{y}$. 

Now, we can write%
\begin{equation}
\mathbf{k}_{c}\left(  t\right)  \equiv k_{p}\left(  t\right)  \widehat
{\mathbf{e}}_{p}+k_{s}\left(  t\right)  \widehat{\mathbf{e}}_{x},
\end{equation}
where%
\begin{align}
k_{p}\left(  t\right)   &  \equiv-\frac{eE_{tp}\tau\cos\left(  \omega
t-\phi\right)  }{\hbar\sqrt{1+\omega^{2}\tau^{2}}},\\
k_{s}\left(  t\right)   &  \equiv-\frac{eE_{ts}\tau\cos\left(  \omega
t-\phi\right)  }{\hbar\sqrt{1+\omega^{2}\tau^{2}}}.
\end{align}
Since we are only interested in the current density in the direction of the
probe field, we only need the $x$-component of the current density. Also, since the probe field is much smaller than the pump field, we can
expand the velocity around the point $k_{p}\left(  t\right)  \widehat
{\mathbf{e}}_{p}$. 

Then, the current that arises to first order in the probe field is found to be
\begin{equation}
\label{eq_Jsx}
J_{sx}\left(  t\right)  \simeq -\frac{e\hbar k_{s}\left(  t\right)  n_{e}}{\overline{M}_{eff}\left(
k_{p}\left(  t\right) \widehat{\mathbf{e}}_{p}\right)  } ,
\end{equation}
where $n_{e}=4\pi k_{F}^{2}/\left(  2\pi\right)  ^{2}$
is the carrier density and
\begin{equation}
\frac{1}{\overline{M}_{eff}\left(  \mathbf{k}_{p}\right)  }\equiv\frac{1}{\pi
k_{F}^{2}}
{\displaystyle\int}
d^{2}\mathbf{k}\frac{1}{M_{eff}\left(  \mathbf{k+k}_{p}\right)  }\Theta\left[
k_{F}-\left\vert \mathbf{k}\right\vert \right]
\end{equation}
is the average inverse effective mass of the carriers over the displaced Fermi disk, where
\begin{equation}
\frac{1}{M_{eff}\left(  \mathbf{k}\right)  }\equiv\frac{1}{\hbar^{2}}%
\frac{\partial^{2}E_{c}\left(  \mathbf{k}\right)  }{\partial k_{x}^{2}}%
\end{equation}
is the inverse effective mass in the $x$-direction at $\mathbf{k}$.\\

The dispersion for
bilayer graphene is given by%
\begin{equation}
\label{eq_Ecdisp}
E_{c}\left(  \mathbf{k}\right)  =\frac{\sqrt{\alpha\left(  k_{x}^{2}+k_{y}%
^{2}\right)  +t_{\bot}^{2}}-t_{\bot}}{2},
\end{equation}
where $\alpha\equiv4\hbar^{2}v_{F}^{2}$
and $t_{\bot}$ is the interlayer hopping energy, which is taken to be \SI{300}{\milli\electronvolt}. 
Using this, we obtain for the $k$-dependent effective mass (\emph{not} the
averaged one) in the $x$-direction:%
\begin{equation}
M_{eff}^{-1}\left(  \mathbf{k}\right) =\frac{1}{m_{o}^{\ast}}\frac{\widetilde{k}_{y}^{2}+1}{\left[  \left(
\widetilde{k}_{x}^{2}+\widetilde{k}_{y}^{2}\right)  +1\right]  ^{3/2}},
\end{equation}
where \begin{equation}
m_{o}^{\ast}\equiv\frac{t_{\bot}}{2v_{F}^{2}}\simeq0.026m_{e}%
\end{equation}
is the effective mass of bilayer graphene at the origin and $\widetilde{k}_{i}\equiv k_{i}/k_{S}$, where
 $k_{S}\equiv {t_\bot}/\sqrt{\alpha}$.\\

Using the expression for the probe current density, given in Eq. (\ref{eq_Jsx}), we can calculate the transmitted probe field at the graphene by using the following equation determined from the field boundary conditions at the graphene:
\begin{equation}
\label{eq_Etrans}
E_{ts}(t) = \frac{2 E_{is}(t)- Z_\circ J_{sx} [ E_{ts}(t) ] }{1+n},
\end{equation}
where $E_{is}(t)$ is the incident probe terahertz field, $Z_\circ$ is the impedance of free space, and $J_{sx} [ E_{ts}(t) ]$ is the probe current density calculated using the transmitted probe field as the driving field. As an approximation in the above expression, for the driving field of the current we use $E_{ts}^{o}(t) = 2 E_{is}(t)/(1+n)$. Using the resulting expressions for the transmitted field with and without the pump, we obtain the following expression for the differential transmitted pump-probe power:
\begin{align}
\begin{split}
\label{eq:TranmissionMeff}
\Delta T  \simeq & \Delta T_{o}\frac{\omega}{2\pi}%
{\displaystyle\int\limits_{0}^{2\pi/\omega}}
\cos\left(  \omega t\right)  \cos\left(  \omega t-\phi\right) \\
& \left\{
\frac{1}{\overline{M}_{eff}\left(  \mathbf{0}\right)  }-\frac{1}{\overline
{M}_{eff}\left(  k_{p}\left(  t\right)  \widehat{\mathbf{e}}_{p}\right)
}\right\}  dt, \\
\end{split}
\end{align}

where
\begin{equation}
\Delta T_{o}\equiv\frac{4Z_{o}\tau e^{2}n_{e}\cos\left(  \phi\right)
}{\left(  1+n\right)  m_{o}^{\ast}\sqrt{1+\omega^{2}\tau^{2}}}.
\end{equation}
Thus, we see that the key quantity of interest is the time average of the inverse
effective mass as a function of $\mathbf{k}$.

\begin{figure}

\subfloat[\label{Band3D}]{
   \centering
   \includegraphics[width=0.7\columnwidth]{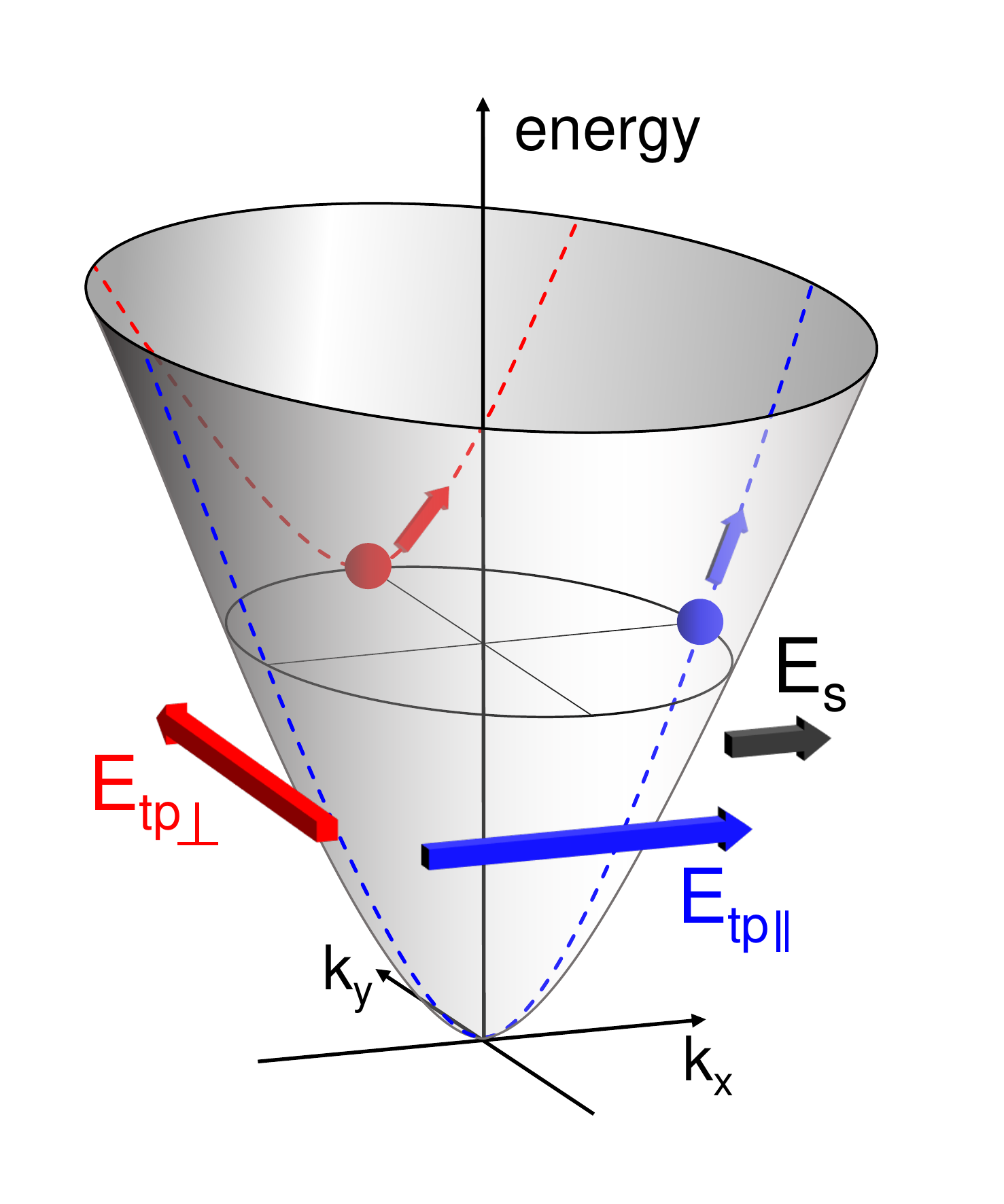}
   }
\hfill
\subfloat[\label{Fig2b_neu}]{
\centering
   \includegraphics[width=0.65\columnwidth]{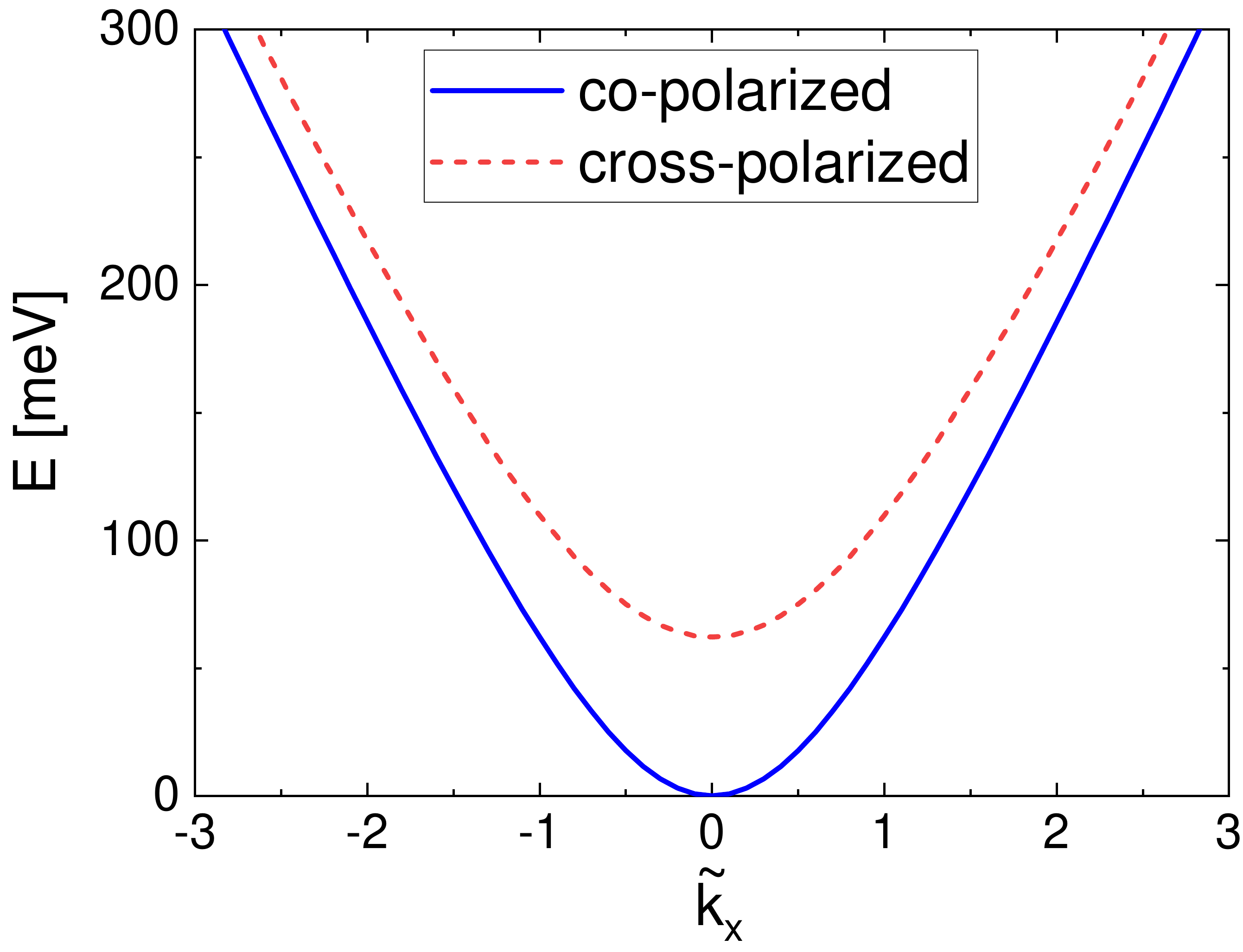}
   }
\hfill
\subfloat[\label{fig_effmass}]{
   \centering
  \includegraphics[width=0.75\columnwidth]{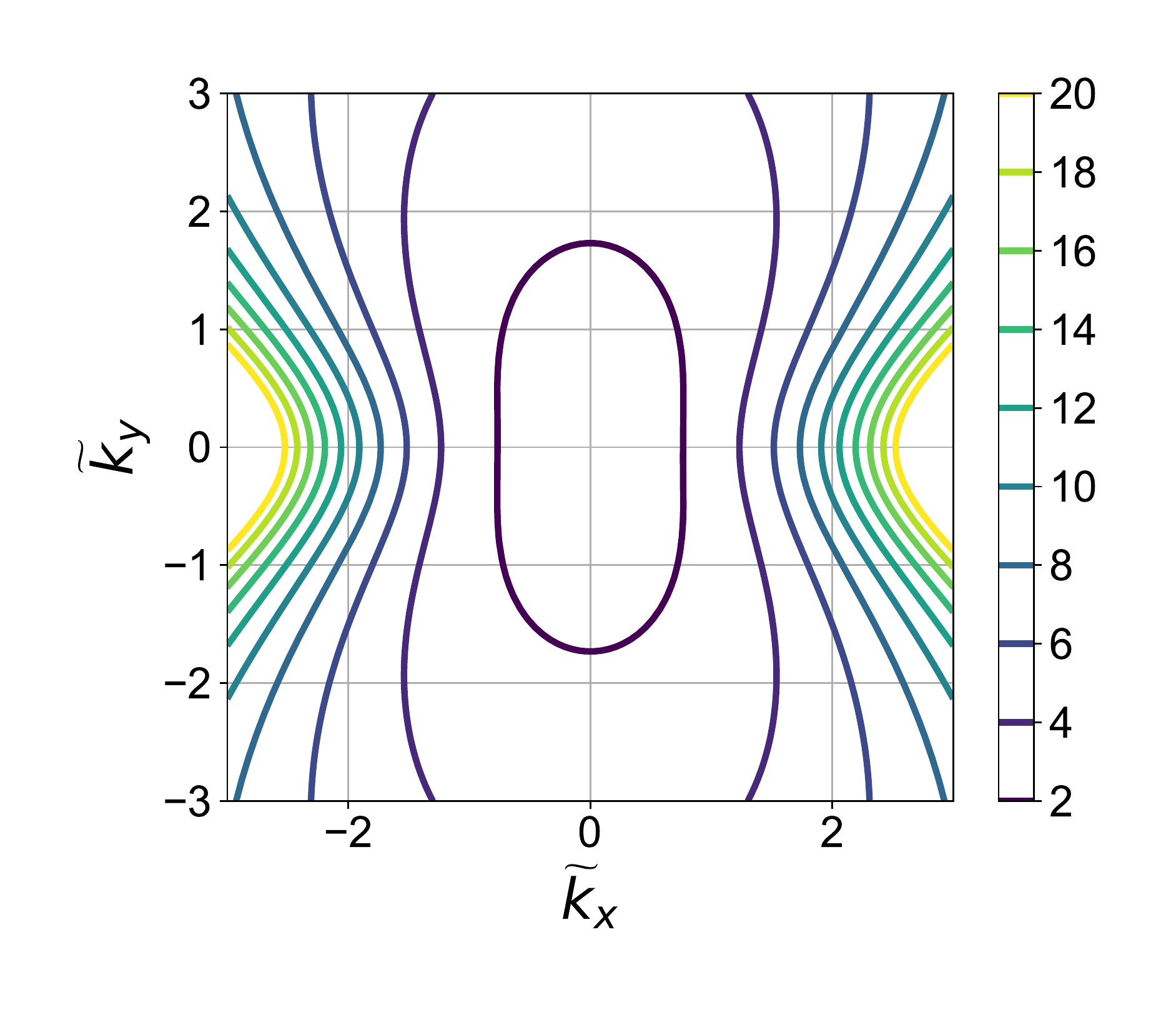}
  }
  \caption{ (a) Conduction band of bilayer graphene. The signal is detected in $x$-direction (black arrow). For co-polarized configuration (blue), the band curvature in $x$-direction is larger than for cross-polarized configuration (red). (b) Band structures in x-direction for constant $k_y$. For co-polarized configuration, $k_y = 0$, for cross-polarized configuration, $k_y = 1$. (c) Contour plot of the effective mass in the $x$ direction in $k$-space relative to the $K$-point. The effective mass is relative to $m_o^*$, such that it is 1 at the origin.}
  \label{fig_2}
\end{figure}

 The $k$-dependence of the x-component of the effective mass is the key to the pump-induced anisotropy. An instructive picture for understanding of this effect is given in Fig. \ref{Band3D}. We are interested in the effective mass in the direction of the probe electric field indicated by the black arrow. Excited electrons at the same absolute value of the pump electric field are depicted for two perpendicular pump fields. In Fig. \ref{Fig2b_neu} we present the band dispersion in the $x$-direction for co-polarized ($k_y = 0$) and cross-polarized configuration (exemplarily choosing $k_y = 1$). In Fig. \ref{fig_effmass}, we present a contour map of the $x$-component of the effective mass in $k$-space.  As can be seen, the effective mass is smallest at the origin.  It increases as one moves away from the origin in either the $x$ or $y$ directions.  However, it increases much more rapidly in the $x$ direction. The transmission scales with the inverse effective mass. Because the effective mass increases much more when the electrons are driven by the pump field in $x$ direction (co-polarization) than when the are driven in the $y$ direction (cross-polarization), the differential transmission is larger in the co- than in the cross-polarized case.

Now, the \emph{average}
effective mass when there is \emph{no pump field present} is given by%
\begin{equation}
\overline{M}_{eff}^{-1}\left(  \mathbf{0}\right)  =\frac{1}{m_{o}^{\ast}}%
\frac{1}{\sqrt{\widetilde{k}_{F}^{2}+1}}.
\end{equation}
Thus, as expected, the average carrier mass increases as the Fermi energy increases. Note that with a Fermi energy of \SI{178}{\milli\electronvolt}, $\widetilde{k}_{F}\simeq 1.93$. There is no analytic result for the average
effective mass or for the average of the differences in the effective masses
when the Fermi disk is displaced along either the $k_x$ or $k_y$ axes. Instead, we
perform a Taylor series expansion in $\widetilde{k}_{p}$ of the mass
difference and then integrate over the Fermi disk, term-by-term. Doing this for the \textit{parallel-pump case} we obtain
\begin{align}
\begin{split}
\left[  \frac{1}{\overline{M}_{eff}\left(  \mathbf{0}\right)  }-\frac
{1}{\overline{M}_{eff}\left(  k_{p}\left(  t\right)  \widehat{\mathbf{e}}%
_{x}\right)  }\right]  & = \\ 
\frac{1}{m_{o}^{\ast}}\left[  3C_{2}\widetilde{k}%
_{p}^{2}-5C_{4}\widetilde{k}_{p}^{4}+7C_{6}\widetilde{k}_{p}^{6}\right], \\
\end{split}
\end{align}
while for the \textit{perpendicular-pump case} we obtain
\begin{equation}
\left[  \frac{1}{\overline{M}_{eff}\left(  \mathbf{0}\right)  }-\frac
{1}{\overline{M}_{eff}\left(  k_{p}\widehat{\mathbf{e}}_{y}\right)  }\right]
=\frac{1}{m_{o}^{\ast}}\left[  C_{2}\widetilde{k}_{p}^{2}-C_{4}\widetilde
{k}_{p}^{4}+C_{6}\widetilde{k}_{p}^{6}\right]  ,
\end{equation}
where%
\begin{align}
C_{2} &  =\frac{4+\widetilde{k}_{F}^{2}}{16\left[  \widetilde{k}_{F}%
^{2}+1\right]  ^{5/2}},\\
C_{4} &  =\frac{24-12\widetilde{k}_{F}^{2}-\widetilde{k}_{F}^{4}}{128\left[
\widetilde{k}_{F}^{2}+1\right]  ^{9/2}},\\
C_{6} &  =\frac{64-144\widetilde{k}_{F}^{2}+24\widetilde{k}_{F}^{4}%
+\widetilde{k}_{F}^{6}}{2048\left[  \widetilde{k}_{F}^{2}+1\right]  ^{13/2}}.
\end{align}
Note first that in the limit that $\widetilde{k}_{p}\rightarrow0$, we obtain%
\begin{equation}
\frac{\Delta T_{\Vert}}{\Delta T_{\bot}}=3.
\end{equation}
Thus, as found in the experiments, the differential signal is expected to be considerably larger for the co-polarized pump configuration. This indicates that this effect is due to the dependence of the effective mass in the $x$-direction on $\mathbf{k}$. \\ 

Now, to proceed, we still need to take the time averages in the expression for the
differential signal.  Doing this, the differential pump-probe power signal for the co- and cross-polarized pump configurations are found to be given respectively by
\begin{align}
\Delta T_{\Vert}  & \simeq\Delta T_{o}\left[  3D_{2}C_{2}\widetilde{k}%
_{po}^{2}-5D_{4}C_{4}\widetilde{k}_{po}^{4}+7D_{6}C_{6}\widetilde{k}_{po}%
^{6}\right]  ,  \label{eq:Tp_HH} \\
\Delta T_{\bot}  & \simeq\Delta T_{o}\left[  D_{2}C_{2}\widetilde{k}_{po}%
^{2}-D_{4}C_{4}\widetilde{k}_{po}^{4}+D_{6}C_{6}\widetilde{k}_{po}^{6}\right], \label{eq:Ts_HH}
\end{align}
where $D_2=3/8$, $D_4=5/16$, $D_6=35/128$, and
$\widetilde{k}_{po}\equiv E_{ip}/E_{S}$, where
$E_{ip}$ is the incident pump field and
\begin{equation}
E_{S}\equiv\frac{t_{\bot}\left(  n+1\right)  \sqrt{1+\omega^{2}\tau^{2}}%
}{4ev_{F}\tau}%
\end{equation}
is a scale field. For scattering time of $\tau=\SI{50}{\femto\second}$, a central frequency of
\SI{3.4}{THz}, and a substrate index of $n=3$, $E_{S}\simeq\SI{89}{\kilo\volt\per\centi\metre}$.
Note that this scaling parameter is relatively sensitive to the scattering time, which is one reason why we find a rather strong dependence on the scattering time in the full simulations that we will present next. For our maximum incident pump
field of, $E_{ip}=\SI{60}{\kilo\volt\per\centi\metre}$, we obtain $\widetilde{k}_{po}\simeq0.17$ (using the transmitted, rather than incident pump field),
so we are in the relatively small-$k_{p}$ regime, even at the highest
fields. This is why in this model the anisotropy factor is essentially 3 for all field amplitudes. Note, however, that for longer scattering times and higher fields, the fifth and seventh order contributions are expected to play a significant role. This will make the anisotropy factor field-dependent and will make the field dependence of the differential transmission deviate from a simple quadratic.  \\
\subsection{Boltzmann equation simulations}
For a quantitative comparison with the experiment, we have performed simulations using  a density matrix formalism in k-space. The dynamic equation for the reduced density matrix elements for carriers in the conduction band is given by the Boltzmann equation \cite{mcgouran2016nonlinear, Marc_Luke_2019effect},
\begin{equation}
\label{eq_Boltz2}
\frac{d\rho_{cc}(\textbf{k},t)}{dt} = 
\frac{e\textbf{E}_t(t)}{\hbar}\cdot \boldsymbol{\nabla
}_{\mathbf{k}}\rho_{cc}(\textbf{k},t) -\frac{[\rho_{cc}(\textbf{k},t) -\rho_{cc}(\textbf{k},0))]}{\tau}, 
\end{equation}
where $\mathbf{E}_t(t)$ is the transmitted THz field (calculated self-consistently), $\rho_{cc}(\textbf{k},0)=f(E_{c}\left(  \vert \mathbf{k}\vert\right)$, where $f(E)$ is the Fermi-Dirac distribution at the initial temperature and chemical potential, and $\tau$ is a phenomenological scattering time that accounts for various scattering mechanisms such as neutral impurities, acoustic and optical phonons, and substrate charged impurities. We take this scattering time to be \SI{50}{\femto\second} (corresponding to $\omega \tau = 1.1$), independent of $\mathbf{k}$. This value gives the best agreement with the experiment and is in accordance with previous studies \cite{Rouhi2012,Hong2009}{}. \\

We use the calculated time-dependent density matrix in Eq. (\ref{eq_current}) to obtain the current density.  This is then used in Eq. (\ref{eq_Etrans}) to obtain the transmitted THz field. The differential transmission signals for the two different polarizations are calculated by simulating the transmission with only the pump or probe, and with both together, with no time delay; the details of the simulation method, the parameters used, and the sensitivity to scattering time and chemical potential are given in Appendix B.

\FloatBarrier


\begin{figure}
 \centering

   \subfloat[\label{fig_3_a}]{
   \centering
   \includegraphics[width=\columnwidth]{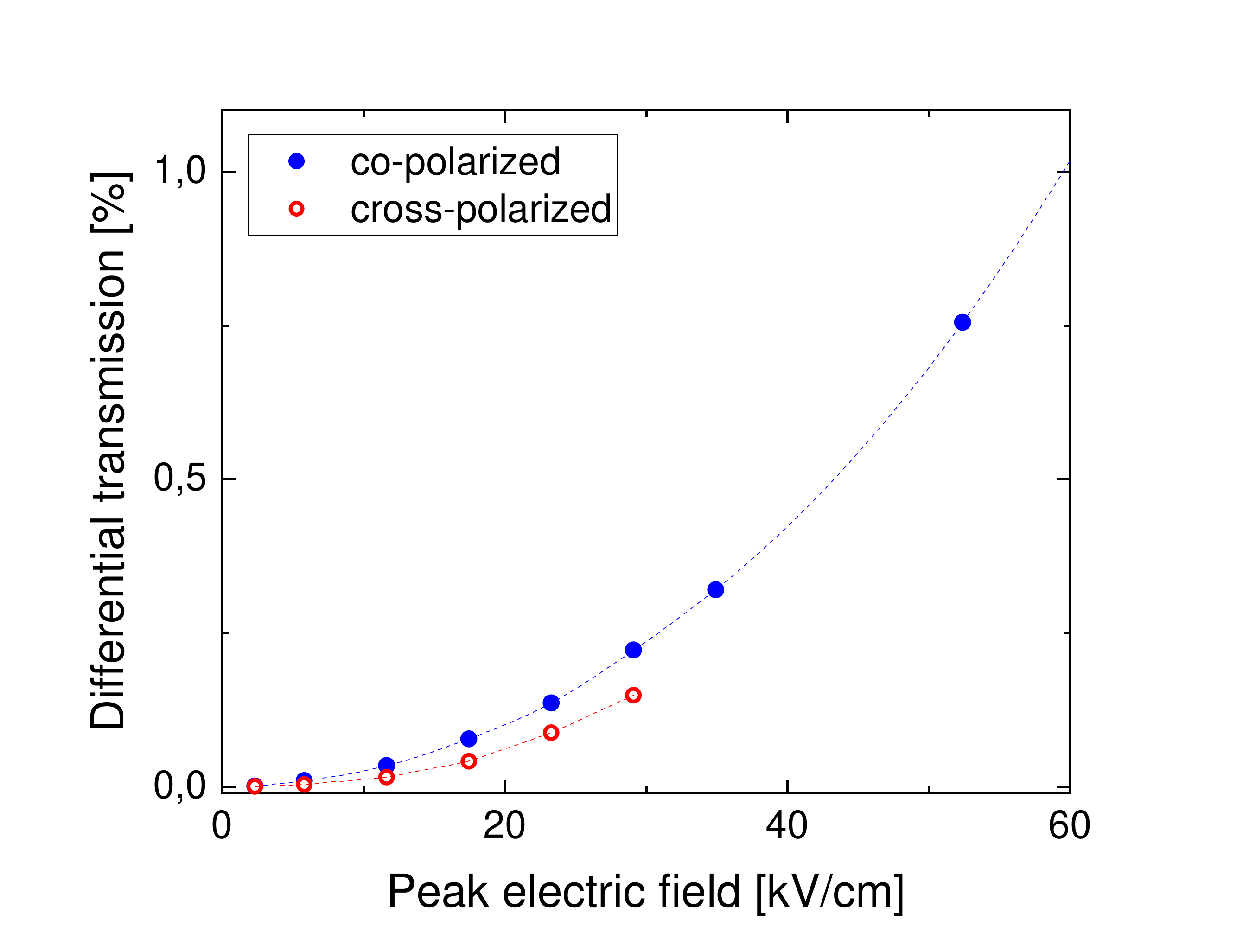}
   }
\hfill
  \subfloat[\label{fig_3_b}]{
   \centering
   \includegraphics[width=\columnwidth]{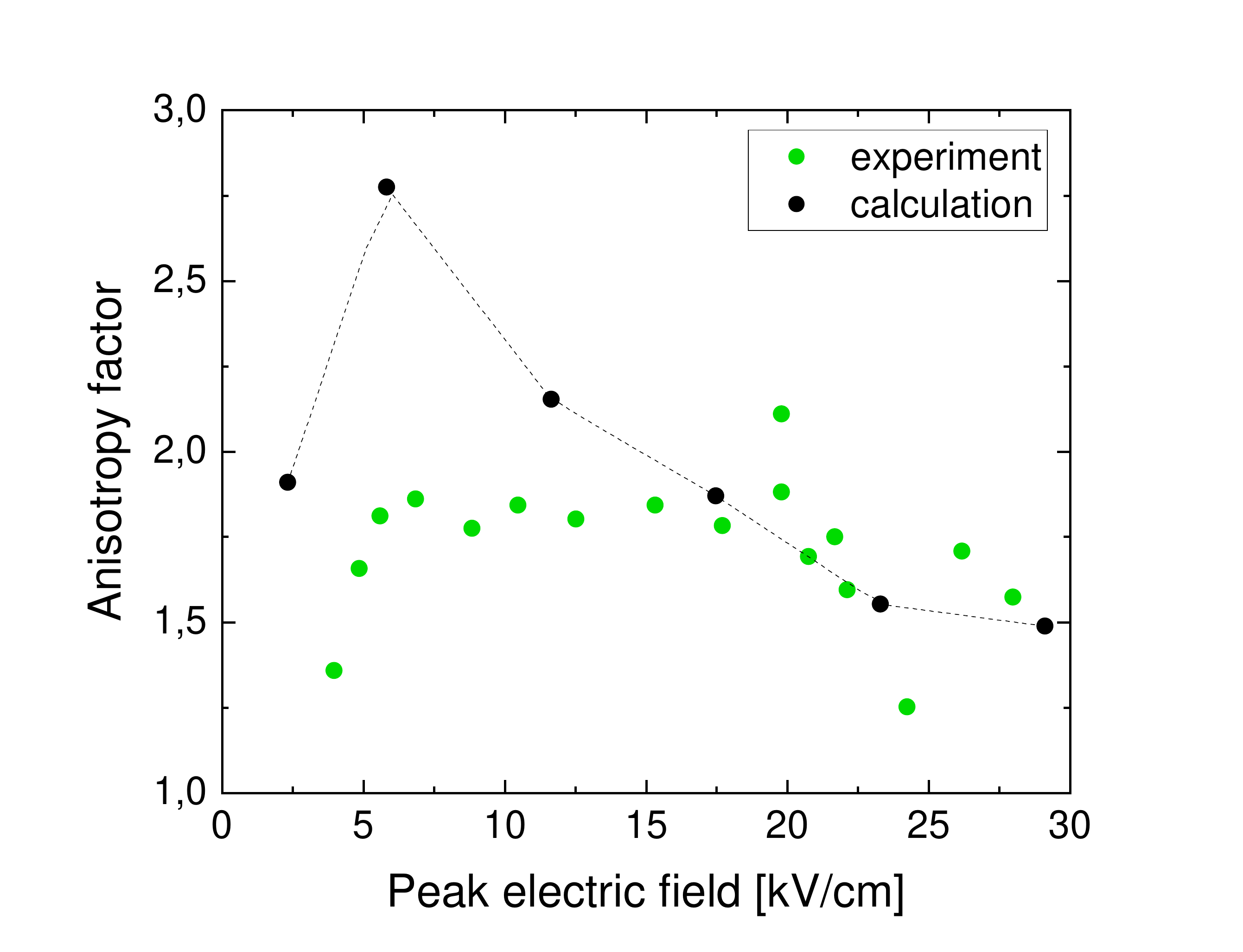}
   }
 
\caption{(a) Simulation results for the differential transmission signal at zero probe delay for co- and cross-polarized pump configurations as a function of the incident THz field amplitude. (b) The anisotropy factor as a function of the incident THz field amplitude found from experiment and simulation. For the simulation results, the dots are calculated values, whereas the dotted lines are interpolations serving as a guide for the eye.}
  \label{fig_3}
\end{figure} 

 In Fig. \ref{fig_3_a}, the calculated differential power transmission for the two different pump polarizations is plotted as a function of the pump peak electric field. The signal amplitudes are in the same range and show a similar anisotropic behavior to that found in the experiment. However, the simulations yield a peak differential transmission for the co-polarized pump configuration of only 1.0\% at the pump field amplitude of \SI{60}{\kilo\volt\per\centi\metre}, while the experiment gives a value of approximately 2.0\%. Also, the dependence of the differential transmission on field is more linear in the experiment than in our simulations. From the semiclassical theory, we find that a seemingly more linear dependence can arise if the nonlinearity is stronger, due to the fifth and seventh order contributions to the response. As has been shown in previous work \cite{Hafez2018,Marc_Luke_2019effect}, a stronger nonlinearity can be obtained if the energy dependence of the scattering time is taken into account.\\
 
In Fig.  \ref{fig_3_b}, we present the experimental and simulations results for the ratio of the co-polarized differential transmission to the cross-polarized transmission, i.e., the anisotropy factor. The agreement is very good at higher field amplitudes, as is the general trend with field amplitude. We find that the average ratio of the co- and cross-polarized induced transmission signals is approximately 1.95 in the simulation, which is close to the experimental value of approximately 1.8. \\ 

In contrast, the simple semiclassical model of the previous section predicted an anisotropy factor that is very close to 3 over the range of field amplitudes considered here. The discrepancy largely arises from the simplifying assumption in the semiclassical theory that even in the presence of scattering, all of the electrons move as a uniform disk in k-space.  Although this is true when there is no scattering, it is certainly not generally true, as carriers will scatter both elastically and inelastically, leading to a more complicated time-dependent distribution. \\

There are a number of factors that likely lead to the differences between the experimental and simulation results. These include uncertainties in the carrier density, the scattering time and the pulse amplitude and shape. As is shown in Appendix C, the differential transmission signal found from our simulations is quite sensitive to the scattering time. In addition, in our simulations we employed an energy-independent phenomenological scattering time. To obtain better quantitative agreement, one should include microscopic scattering due to neutral and charged impurities, phonons and electron-electron interactions. We have included the effects of neutral impurity and optical phonon scattering on the nonlinear THz transmission of monolayer graphene in previous work \cite{Marc_Luke_2019effect} and found that this generally leads to a stronger nonlinearity relative to simulations with a constant scattering time. We plan to include these effects and others on the differential transmission of bilayer graphene in future work. \\

\FloatBarrier
\section{Further discussion}

Finally, we examine our results in the context of previous related work. The concept of effective mass anisotropy was used in a THz pump-THz probe study on InGaAs. In the experiment, an anisotropy factor of 1.5 was obtained \cite{Blanchard2011}. The anisotropy of the optical intraband excitation was found to be nearly independent of the Coulomb and phonon scattering processes. It was attributed solely to the different energy band curvatures, i.e. different effective masses in different $k$-directions \cite{Blanchard2011}{}. In contrast to InGaAs, bilayer graphene is a purely 2D system. We used pump fields that are three times lower, and we obtain a significantly larger anisotropy factor. In addition, we investigated the dependence on the pump electric field.\\

We note that an anisotropic response has also been observed in pump-probe experiments studying the \emph{interband} response in graphene \cite{Mittendorff2014,Trushin2015,KoenigOtto2016}{}. In those experiments the induced transmission in co-polarized configuration was more than twice as large as compared to the cross-polarized case. The underlying physics of the interband excitation, however, differs strongly from the case of intraband excitation. Due to selection rules, the interband excitation results in stronger occupation of states in the direction perpendicular to the orientation of the field, exactly the opposite of the situation for intraband excitation. In interband excitation experiments with photon energies below the optical phonon energies, the anisotropy factor decreased strongly with increasing pump fluence. This is attributed to non-collinear Coulomb scattering, which scales with the number of carriers. In our intraband excitation experiment, no extra carriers are generated and the anisotropy factor is nearly constant. The situation is different for the first observation of the anisotropy in Ref. \cite{Melnikov2019}{}. In that THz pump – optical probe experiment, the anisotropy factor is fluence-dependent. Furthermore, the anisotropic response is five orders of magnitude lower than what we obtained in our experiment. Optical phonons may play an important role in that case in redistributing the carriers that are then excited by the \SI{800}{\nano\metre} probe. This would explain the very weak anisotropic response relative to what we find. \\

\section{Conclusion}

In conclusion, we performed degenerate THz pump-probe experiments on bilayer graphene with linearly polarized radiation. We observe an almost linear dependence of the pump-probe signal on the electric field in both co- and cross-polarized configurations and a nearly field-independent anisotropy factor. We find from both a simple semiclassical model and a more complete numerical simulation of the Boltzmann equation that the source of the anisotropy is the dependence of the diagonal component of the effective mass in bilayer graphene on the $k$-vector. In addition, we show that although scattering is not the primary source of the nonlinear effects, the results are quite sensitive to the scattering time. Using a Boltzmann simulation, we obtain good agreement with the experimentally observed anisotropy factor for a scattering time of \SI{50}{\femto\second}. In future work, we plan to include energy-dependent scattering at a microscopic level, which we believe will yield even better agreement with the experimentally observed field dependence of the differential transmission signal. Our findings may open new application perspectives, e.g. THz harmonic generation in a broad frequency range directly related to the nonparabolicity of the bands.\\



\section{ACKNOWLEDGEMENT}

A. Seidl, H. Schneider, M. Helm and S. Winnerl thank P. Michel and the FELBE team for their dedicated support. M.M. Dignam and R. Anvari thank the Natural Sciences and Engineering Research Council of Canada (NSERC) and the Canadian Foundation for Innovation (CFI) for funding as well as Compute Canada for computational resources.

\appendix

\FloatBarrier
\label{Appendix A}
\section{Additional experimental results}

\begin{figure}[ht]
 \centering
    \includegraphics[width=\columnwidth]{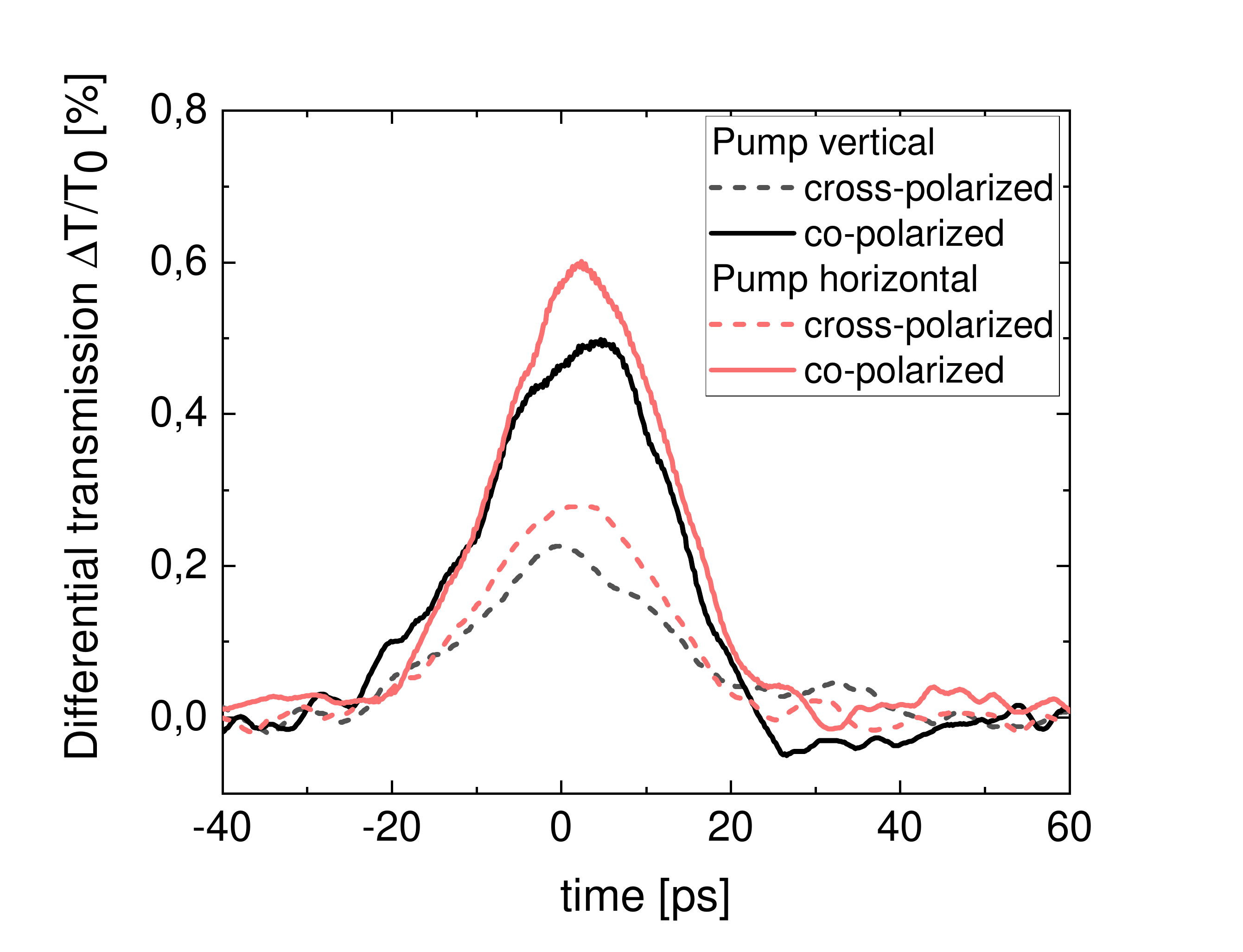}
\caption{Differential transmission signal as a function of probe delay for all four combinations of horizontally and vertically polarized pump and probe beams at \SI{2}{THz} for a pump amplitude of \SI{3}{\kilo\volt\per\centi\metre}. Here, the anisotropy factor is 2.1 for vertical and 2.2 for horizontal pump polarization.}
  \label{fig_4}
\end{figure}

To exclude effects of the sample orientation and the setup, a control experiment was performed at \SI{2}{THz}. There, we measured the pump-probe signal in all four combinations of horizontally and vertically polarized pump and probe beams. As shown in Fig. \ref{fig_4}, the absolute transmission changes by about \SI{20}{\percent} for the different pump polarizations but the anisotropy factor stays the same within the measurement uncertainty.

\FloatBarrier
\label{Appendix B}
\section{Simulation details}

In this Appendix, we present some of the details of the calculations used to evaluate the differential signals using our full numerical simulation. \\

To solve the Boltzmann equation (Eq. (\ref{eq_Boltz2})), a finite difference approximation to the gradients is applied and a fourth-order Runge-Kutta method is used to solve the resulting equations on a hexagonal grid uniformly sampled in k-space about the $K$ and $K'$ points.  The current density as a function of time is given by Eq. (\ref{eq_current}) in the main document, from which we calculate the transmitted field. Note that the field that drives the carriers is the self-consistent transmitted field at the graphene, not the incident field. To obtain the pump-probe signals for the two different pump configurations, we perform the simulations of the transmitted signal both with and without the pump present (see below). \\

The input terahertz pump and probe field are linearly polarized sinusoidal Gaussian pulses with a central frequency of \SI{3.4}{THz}. We take the pulse duration to be \SI{3.5}{\pico\second} (FWHM). This is considerably shorter than the experimental duration of \SI{13}{\pico\second}, but we find that our results do not change much when we increase the duration, and longer pulses increase the computational time significantly. We take the peak value of the incident field of the probe to be \SI{100}{\volt\per\centi\metre}, which is high enough to avoid numerical noise and low enough to avoid nonlinear effects due to the probe field alone.  The bandstructure used in the simulation is that given in Eq. (\ref{eq_Ecdisp}) and the chemical potential is set to \SI{178}{\milli\electronvolt}, such that the carrier density is \SI{6.5e12}{\per\square\centi\metre}. The index of refraction of the substrate is taken to be $n=3$, which is the refractive index of SiC at terahertz frequencies. Finally, we take the temperature to be \SI{12}{K} and we choose a phenomenological scattering time of \SI{50}{\femto\second} (corresponding to $\omega \tau = 1.1$), as discussed in the main text. Although we use an energy-independent scattering time, the form of our dynamic equation is such that it includes both elastic and inelastic scattering, with the electron density always relaxing back to thermal equilibrium. In this way, it differs from our simple semiclassical model. Below, we examine the sensitivity of the calculated differential transmission to the chemical potential and scattering time. \\

We now turn to the calculation of the pump-probe signals. The co-planar transmitted energy is defined as the transmitted energy of a field that consists of the difference between the transmitted field when both pump and probe fields are present ($\mathbf{E}_{t}^{pu+pr}$), and the transmitted field when only pump field is present ($\mathbf{E}_t^{pu}$), normalized to the incident energy of the probe ($\mathbf{E}_i^{pr}$)

\begin{equation}
\label{eq_2}
T^{\parallel} = \frac{\int dt \, ( \mathbf{E}_{t}^{pu+pr}\cdot \widehat{\mathbf{e}}_{y} - \mathbf{E}_t^{pu}\cdot \widehat{\mathbf{e}}_{y} )^2 }{\int dt \,  (\mathbf{E}_i^{pr} \cdot \widehat{\mathbf{e}}_{y})^2 }.
\end{equation}

Here, all signals are chosen to be polarized in the $\widehat{\mathbf{e}}_{y}$-direction (it was discussed in the main text that the differential transmission signals are isotropic), which is emphasized by performing the dot product in the above equation.

The cross-transmitted energy when pump and probe are in the $\hat{x}$ and $\hat{y}$ directions, respectively is given by
\begin{equation}
\label{eq_3}
T^{\perp} = \frac{\int dt \, ( \mathbf{E}_{t}^{pu+pr}\cdot  \widehat{\mathbf{e}}_{y} )^2}{\int dt \,  (\mathbf{E}_i^{pr} \cdot \widehat{\mathbf{e}}_{y})^2 },
\end{equation}
where we have calculated the transmitted energy in the $\hat{y}$ direction (assuming that the nonlinear effect of pump in the transverse direction is negligible within the range of the studied incident pump-fields),  normalized to the incident energy of the  probe signal. Finally, the transmitted energy when only probe is present (always polarized in the $\widehat{\mathbf{e}}_{y}$-direction) is given by

\begin{equation}
\label{eq_4}
T^{\circ} = \frac{\int dt \,  (\mathbf{E}_t^{pr} \cdot \widehat{\mathbf{e}}_{y})^2 }{\int dt \,  (\mathbf{E}_i^{pr} \cdot \widehat{\mathbf{e}}_{y})^2 } \,\, .
\end{equation}

The co-planar and cross differential transmission signals can be simply obtained by
$\Delta T^{\parallel} = (T^{\parallel} - T^{\circ} )/ T^{\circ}$, and $\Delta T^{\perp} = (T^{\perp} - T^{\circ} )/ T^{\circ}$, respectively. \\

\label{Appendix C}
\section{Effect of scattering time and chemical potential on simulation results}

In this Appendix, we examine the effects of the scattering time and chemical potential on the calculated differential transmission. \\

\begin{figure}[ht]
   \centering
 \subfloat[\label{fig_ap_1a}]{
   \includegraphics[width=\columnwidth]{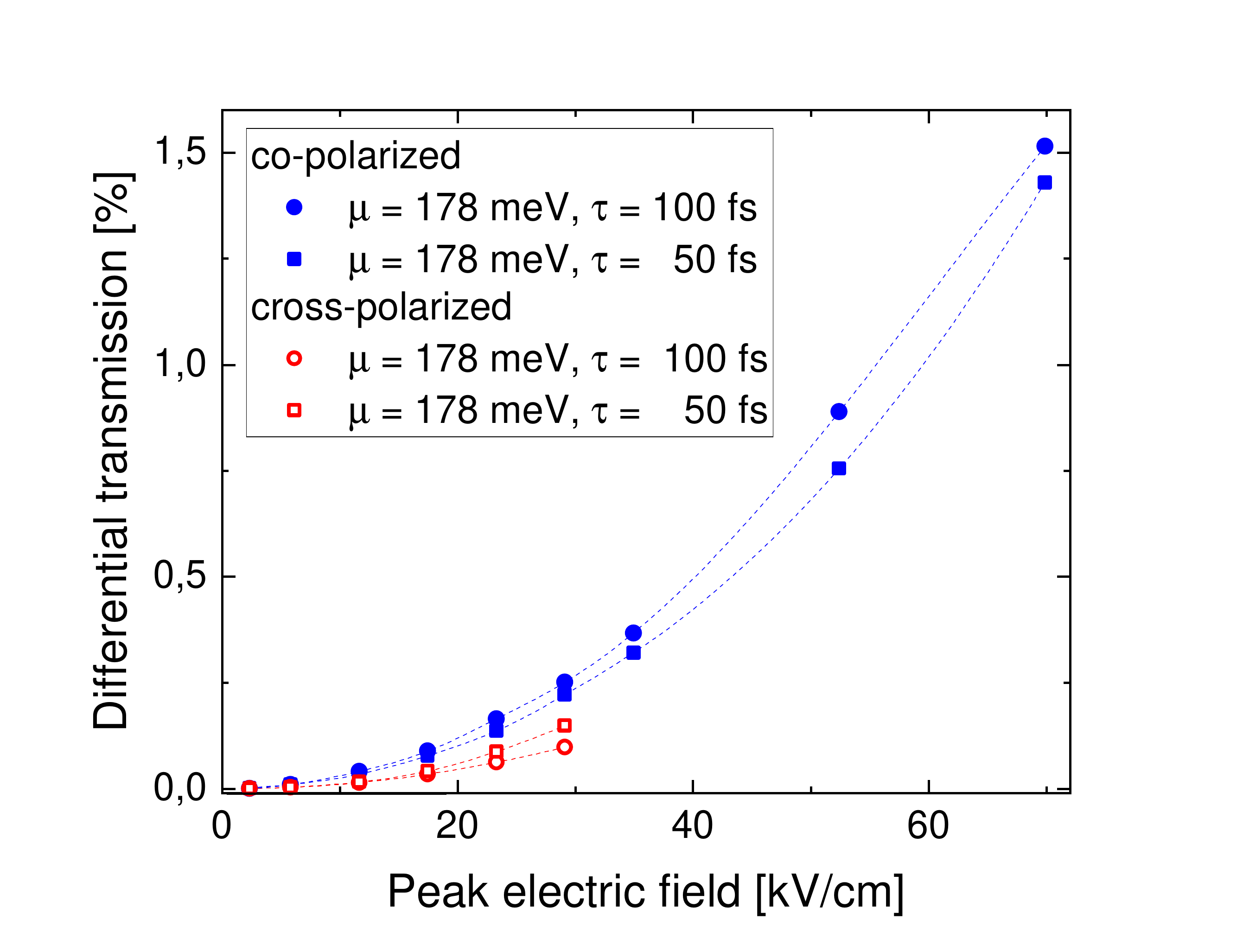}
   }
\hfill
   \centering
    \subfloat[\label{fig_ap_1b}]{
   \includegraphics[width=\columnwidth]{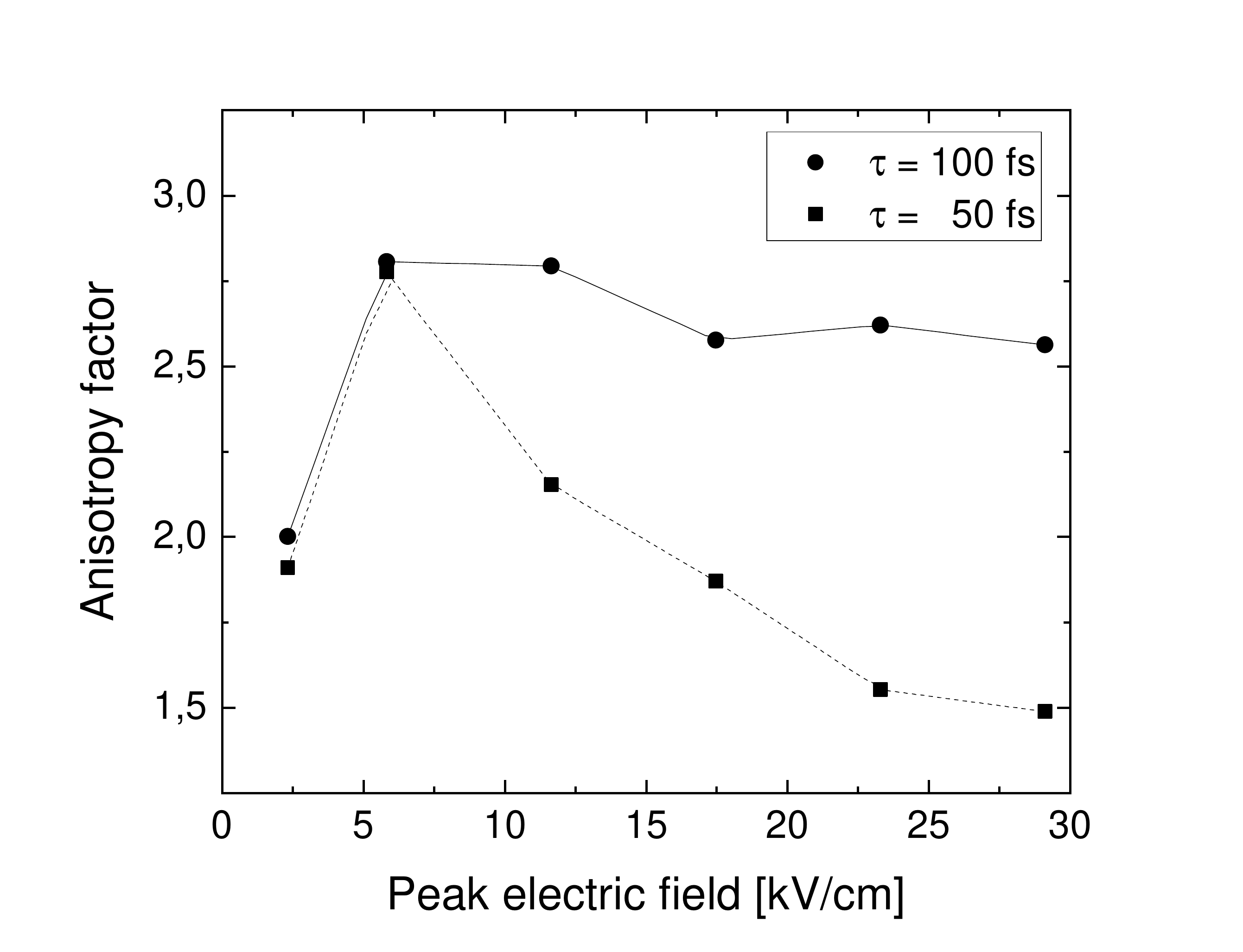}
   }
\caption{Effect of the scattering time on (a) the differential transmission and (b) the anisotropy factor. In (a), the co- and cross-polarized signals are indicated by closed and open symbols, respectively. In (a) and (b), circles are for a scattering time of \SI{100}{\femto\second} and squares are for a scattering time of \SI{50}{\femto\second}. In all cases, the chemical potential is \SI{178}{\milli\electronvolt}.}
  \label{fig_ap_1}
\end{figure} 

Fig. \ref{fig_ap_1} shows the effect of the phenomenological scattering time on the differential transmission and anisotropy. We observe that at lower scattering time the numerically calculated anisotropy factor averaged over incident fields (2.51 and 1.95 for $\tau$ = \SI{100}{\femto\second} and $\tau$ = \SI{50}{\femto\second}, respectively) is closer to the experimental average value of 1.8. \\

\begin{figure}
 \subfloat[\label{fig_ap_2a}]{
   \centering
   \includegraphics[width=\columnwidth]{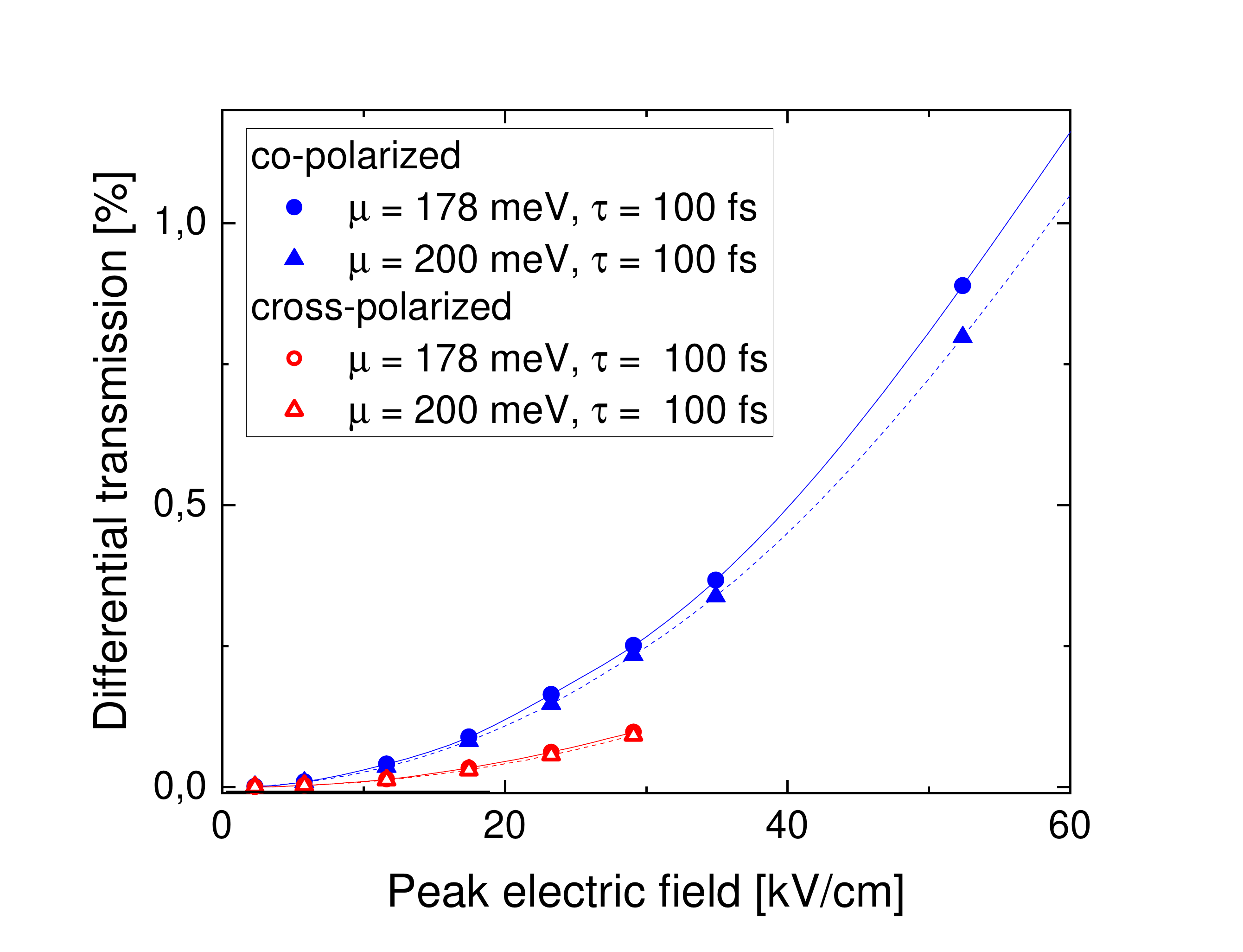}
   }
\hfill
 \subfloat[\label{fig_ap_2b}]{
   \centering
   \includegraphics[width=\columnwidth]{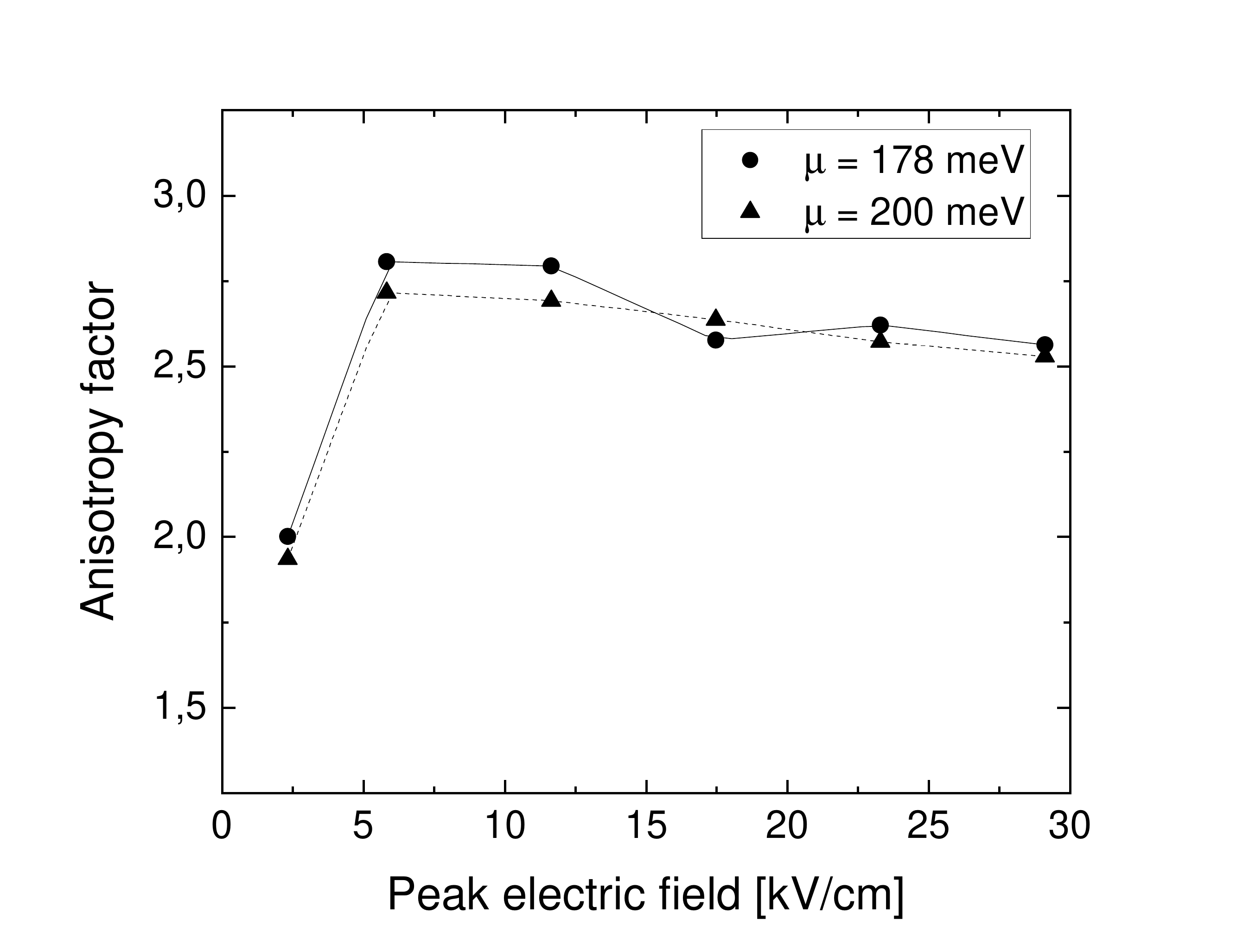}
   }
\caption{Effect of the chemical potential on (a) the differential transmission and (b) the anisotropy factor. In (a), the co- and cross-polarized signals are indicated by closed and open symbols, respectively. In (a) and (b), circles are for a chemical potential of \SI{178}{\milli\electronvolt} and triangles are for a chemical potential of \SI{200}{\milli\electronvolt}. In all cases, the scattering time is \SI{100}{\femto\second}.}
  \label{fig_ap_2}
\end{figure}

Fig. \ref{fig_ap_2} shows the effect of the chemical potential on the differential transmission and anisotropy factor. Here we observe that the difference between co- and cross-polarized transmission signals reduces as the chemical potential increases, and hence causes the anisotropy of the differential transmission signal to decrease.

\FloatBarrier

%

\bibliographystyle{apsrev4-2}
\bibliography{thebibliography}
\end{document}